\documentclass[twocolumn,showpacs,preprintnumbers,amsmath,amssymb]{revtex4}

\begin{document}
\draft

\title{Charge fluctuations in coupled systems: ring coupled to a wire or ring}

\author{P. Singha Deo$^1$, P. Koskinen$^2$, and M. Manninen$^2$}
\address{$^1$ Unit for Nano-science $\&$ Technology,
S. N. Bose National Centre for Basic Sciences, JD Block,
Sector III, Salt Lake City, Kolkata 98, India.\\
$^2$ Nanoscience Center, Department of Physics, 
University of Jyvaskyla, PO Box - 35,
40014 Jyvaskyla, Finland.}
\date{\today}

\begin{abstract}
Coupled systems in mesoscopic regime are of interest as charge fluctuation 
between the sub-systems will depend on electron-electron interactions and will
play a dominant role in determining their thermodynamic properties.
We study some simple systems like a stub or a bubble strongly
coupled to a ring. We show that for strong electron-electron interaction,
there are some regimes where these charge fluctuations
are quenched and charge is individually conserved in the two subsystems.
This feature does not depend on choice of parameters or charge distribution.
\end{abstract}


\maketitle

\section{Introduction}

Coupled mesoscopic samples are very challenging to understand.
Transfer of charge and charge fluctuations between different
parts of a coupled system, can dramatically alter its properties
\cite{sta,ryu}.
Electron-electron interactions strongly dictate such
charge transfer and fluctuations. 

If one starts with a system that has translational
invariance or rotational invariance then
electron-electron interactions preserve the invariance.
This often helps in solving an interacting electron system.
However, if we start with a system that has no translational
or rotational invariance, the the problem of solving the interacting
electrons in the system is non-trivial.
Coupled systems do not have rotational or translational
invariance and this makes the problem very difficult.

In this regard, simple systems like finite one dimensional (1D) systems,
are what researchers intend to understand first \cite{sta,ryu}.
With the advent of mesoscopic systems, one can now realize
such finite and 1D systems in the laboratory and this has further
encouraged scientists to examine the role of interactions
in such simple systems \cite{sta,ryu} as that considered in Fig. 1 (a).
It consists of a 1D ring connected to a 1D finite wire (referred
to as stub). In Fig. 1 (b) we replace the stub by a small ring
that we refer to as bubble. An
Aharonov-Bohm flux through the center of the ring induces a 
persistent current in the ring \cite{but}, due to which the system
has a net magnetization. There is no flux through the bubble.
The persistent current in the $n$th energy
level is defined
as $I_n=-c {\partial E_n \over \partial \phi}$, where $E_n$ is
the $n$th energy eigenvalue, $c$ is the velocity of light
and $\phi$ is the flux through the ring.
States in the ring are plane wave like extended states that
carry current (due to the flux piercing it), while the states in the stub 
or in the bubble are like the localized
states of a cavity.  These systems
have been studied earlier by various authors. The parity effect was
studied in \cite{deo1, deo2}, persistent currents in such a system in
the diffusive regime was studied in ref. \cite{pas},
conductance across such a system \cite{cho} and persistent currents
in such a system in the ballistic regime \cite{pek} has also been studied.

Buttiker and Stafford \cite{sta} considers the
stub to be 
weakly coupled to the ring. In this weak coupling limit 
they consider the regime wherein in the non-interacting limit
an empty stub state (ESS) weakly hybridizes with an occupied
ring state (ORS) 
and the Fermi energy coincides with the hybridized states. 
In this regime an effective theory was put forward to account
for electron-electron interactions.
The effective theory ignored the presence of all other states
apart from the ESS and ORS and included electron interactions
with the help of a geometrical capacitance.
On the other hand, Cho et al \cite{ryu}, consider again the weakly
coupled regime, when the stub state is below the Fermi energy,
in the Kondo regime. Electron-electron interactions was treated
in the infinite $U$ limit of the Hubbard model, using the slave boson
representation.

\section{The problem}

On one hand, the quantum mechanical uncertainty tends to create
charge fluctuations in the system and on the other hand the
electron-electron interactions tend to localize the electrons
and destroy the charge fluctuations. Situations, wherein
charge fluctuations are completely quenched are of fundamental
importance in physics. Several novel effects like quantum
phase transitions, deformations of nuclei \cite{mot}, deformations of
electronic states in quantum dots \cite{kos} and clusters \cite{cle},
fractional periodicity of persistent currents \cite{deo0} are all
related to quenching of charge fluctuations. So such a competition
between localization and delocalization may also play some important
role in the case of charge fluctuations in
coupled mesoscopic systems and this is what we intend to
investigate in this paper.

In this paper we consider the regime of strong coupling between
the sub-systems for
any arbitrary interaction strength. In the
strong coupling regime, it is not possible to ignore
states other than ESS and ORS. Also it is not possible to
treat electron-electron interactions in the form of a geometrical
capacitance or the slave boson formalism. 
We solve the problem numerically, making exact 
diagonalization of Hubbard type Hamiltonian.
We show that
at certain Fermi energies, and at certain regimes which we call
the Fano regime, the electronic states of the 
strongly interacting system
can be predicted by looking at the non-interacting single particle levels.
Charge fluctuations between the ring and the stub can be easily
quenched in these special regimes by electron interactions and
this is the main reason for such behavior. There is a non-symmetry
dictated node (NSDN) in the many body wave function in this Fano regime
that is responsible for making charge conserved and integral
in the two sub-systems and hence such a behavior. In other regimes
such a simple picture cannot be visualized.

\section{The model}

Schematic figures of the geometries considered are shown in Fig. 1
(a) and (b).
The systems are made up of sites (shown as bold dots)
to be described by the 
Generalized Hubbard Model. The sites in the
stub or in the bubble are labeled as A and B.
Due to the presence of the stub or bubble, 
the system has no rotational symmetry.
We use the generalized Hubbard model to find the electronic states
of the system. The generalized Hubbard Hamiltonian is
\begin{eqnarray}
H &=& \sum_{i,\sigma} \epsilon_i c_{i,\sigma}^\dagger c_{i,\sigma}
+ \left(te^{i \alpha/R}
\sum_{<ij>,\sigma} 
{c_{i, \sigma}^\dagger c_{j, \sigma}} + HC\right) \nonumber\\
& &+U \sum_{i,\sigma \ne \sigma'} {n_{i, \sigma} n_{i, \sigma '}}
+ V \sum_{<ij>,\sigma} {n_{i, \sigma} n_{j, \sigma'}}
\end{eqnarray}
 Here $n_i=c_i^\dagger c_i$, $c_i$ is electron annihilation operator at site
i, $\sigma$ is spin index, and $\epsilon_i$, $t$, $U$, $V$ are parameters
that have their usual meanings: on-site energy, hopping integral,
on-site Coulomb interaction and nearest neighbor Coulomb interaction,
respectively. $R$ is the total number of sites
in the ring only, $\alpha=2 \pi \phi / \phi_0$, $\phi_0= hc/e$
and $\phi$ is the flux through the ring. 
The number of sites making the stub or bubble is denoted as S.

\section{The single particle states}

In this section we shall briefly review what is known about the
single particle states of coupled systems.
In Fig. 2. we show the single particle levels for U=0 and V=0 for a
system as in Fig. 1 (a),
with 9 sites in the ring and 2 sites in the stub. Thus this
is a non-interacting system and the first 6 single particle levels 
as a function of flux are shown
in Fig. 2. The ground state is diamagnetic as it has a minimum
at zero flux and energy increases with flux. The first excited
state is paramagnetic as it has a maximum at zero flux and energy
decreases with flux. All the 6 energy levels have a magnetization
associated with it except the 4th.
One can see that the 4th level from below (i.e., the
3rd excited state) is independent of flux. This is due to the NSDN
\cite{sre}. There is a node in the wave-function
of the 3rd excited state at the junction between the ring and the
stub. We refer to this node as non-symmetry dictated node, a term
originally coined by Leggett \cite{leg}. Due to this node,
the wave-function in the 3rd excited state do not see the magnetic
flux.  Note that the energy of this state is $E=-E_0$.
There can be several symmetry dictated nodes (like nodes
due to box quantization or nodes due to antisymmetry property
of many body wave-function) that do not make a state flux
independent. The NSDN is forced by the boundary condition
at the free end of the stub.
If we cut the ring at one point and attach leads to the two
broken ends then we get the system shown in Fig. 3 (a) and (b).
The transmission
through this system at an energy $E=-E_0$ show an absolute 0
due to this NSDN. Such a resonance is called Fano resonance \cite{tek}.
It is easy to find situations wherein such flux
independent states occur as it can be determined from diagonalization
of the non-interacting Hamiltonian. 
Alternately, it can be found even more easily, by inspecting
the numbers of sites
and this follows from a general theory of parity effect \cite{sre}.
For example it happens for
a (6+2) system, (12+2) system, (15+2) system, (9+3) system and so on.
By (6+2) system we mean 6 sites in the ring and 2 sites in the stub
and so on. A (9+2) system will have eleven levels that will
change slope 9 times as there are 9 sites in the ring.
The ground state will be diamagnetic. The first excited state will
change slope for the first time to be paramagnetic. To change the
slope 9 times, the 4th and 8th state has to be flux independent
or else the particle hole symmetry will be broken.
Even if we place the two sites of the stub as in Fig 1 (b)
so that the two sites make a bubble, even then the 3rd excited state
is independent of flux. 
We shall show that the strongly interacting system
in situations where single particle levels show flux independence has
some general features that do not depend on parameters of the system.

\section{The many body states}

We take the same system as that in Fig. 2, i.e., a 
ring-stub system with S=2 and R=9,
and put 4 spin up and 4 spin
down electrons in it. In the non-interacting limit, where each
single particle level is 2 fold spin degenerate, the Fermi energy
will coincide with the 4th level which is flux independent. The 4th level
will accommodate 1 spin up electron and 1 spin down electron and the
persistent current or magnetization of the system will depend
on 3 up and 3 down electrons in the first 3 energy levels. A small
amount of electron-electron interaction or disorder in site energies
will perturb the total
energy of the system and one cannot think of any particle
not affected by flux.
In Fig. 4 we plot the occupation
probability $<n_A>$ of site A and $<n_B>$ of site B in the
presence of a small amount of interaction. Here $<n_A>$
means the expectation value of $c_A^\dagger c_A$ and so on.
We have considered $<n_A>$ and $<n_B>$ for spin up electrons
only as that for spin down is identically the same due to
spin degeneracy in absence of an explicit magnetic field.
One can see that the total occupation probability of the
two sites in the stub, i.e., the sum of $<n_A>$ and $<n_B>$
is strongly flux dependent. It means that Aharonov-Bohm flux
can cause charge transfer and charge fluctuation between the
stub and the ring. This is a gauge dependent charge transfer
in a system at equilibrium
as opposed to non-equilibrium
bias created charge transfer. Normally in quantum
dots, one uses the bias induced charge transfer, wherein a small
voltage bias is applied to pump a charge into the quantum dot.
In a closed coherent system as that considered in this work, changes
in equilibrium 
parameters in the Hamiltonian can lead to such charge transfers.
Such charge transfers can drastically alter the thermodynamic
properties of the sample like in this case the magnetization
of the sample. Hence it is important to understand these charge
transfers. Also it is important to understand
situations wherein charge remains conserved, does not fluctuate,
and what kind of conservation it is.
For example we have seen that in the non-interacting limit, the
4th spin up and spin down electrons remain completely conserved
in the stub in a flux independent state. This happens because
of the NSDN (non-symmetry dictated node) in the non-interacting
wave function. 
But once there are interactions, the non-interacting
wave functions are perturbed. Then the NSDN has
little or no role to play and everything is dominated by interactions.

However, if we make $U$ and $V$ much larger, then there is
the same amount of 
integer charge localized in the stub as in the non-interacting
case. This can be
seen in Fig. 5. This is counter-intuitive. Because as we increase
the repulsive interaction between electrons, we will expect that
the charge in the two neighboring sites in the stub will decrease
as long as total charge concentration in the ring is lower than
in the stub. But instead the stub sites acquire more and more
charge in order to make the total charge in the stub same as that
in the non-interacting case and that is an integer. 
Here if we take the sum of $<n_A>$ and $<n_B>$, then the sum is
constant in flux and close to 1 as shown by the dotted line in Fig. 5. 
They can individually change by appreciable
amounts, implying that we have not yet reached the localized regime
wherein each electron gets localized into a crystal \cite{deo0}. 
Charge can
fluctuate between sites $A$ and $B$ and so can charge fluctuate
between different sites in the ring to carry an appreciable amount
of persistent current. But total persistent current is like
that of 3 spin up and 3 spin down electrons in a ring. This will
be illustrated in detail below.

We want to stress that this is a general feature and it happens for
all system sizes provided it is in the Fano regime, which means
that in the non-interacting limit there are flux independent states
that coincide with the Fermi energy.
Even if we put some disorder in the site energies, we get the same
result as in Fig. 5,
as long as the site energies are much smaller than the
strength of electron-electron interactions.
For example if we take a 6+2 system then in the non-interacting
limit, the 3rd single particle state is flux independent. So,
if we put 3 spin up and 3 spin down electrons in the system, then in the
strongly interacting limit, 1 up and 1 down electron remain completely
localized in the stub and the magnetization of the system is
determined by 2 up and 2 down electrons in the ring. 
In Fig. 6, we have taken a 6+2 system with 3 up and 3 down electrons in
it. We have also taken the strong interaction limit, when 1 up
and 1 down electrons are localized in the stub. We have
plotted the first 6 many body energy levels of the system
that exhibit large gaps at the zone boundaries $\alpha=0, 0.5, 1$
and level crossings within the zone. Level crossings within the
zone is a precursor of Wigner crystallization \cite{deo0}
and is not related to the breakdown of rotational symmetry.
The gaps at the zone boundary are due to the absence of rotational
symmetry.
In Fig. 7, we have taken
just a ring of 6 sites with a defect, and plotted the first
6 energy levels as a function of $\alpha$ when there are 2
up and 2 down
electrons in the ring, in the same strong interaction limit as in Fig. 6.
The energy levels are very similar to that in Fig. 6, which
implies that in Fig. 6, the persistent current is completely
determined by 2 up and 2 down electrons in the ring.
Charge localized in the stub does not contribute to persistent current
and also charge
fluctuations inside the stub has no qualitative effect on the
persistent current in the ring. There can be non-interacting
situations \cite{hen}
wherein charge fluctuations do not influence propagating states
and in our system this seems to happen for strong
interactions in the Fano regime.

Note that when the stub is made of only 2 sites, then
in strong interaction limit, accommodating 1 up and 1 down
electrons is almost pushing the stub to its maximum
capacity. If we make
the stub having more sites, then it happens more easily and with smaller
values of $U$ and $V$. For example, if we take a 8+3 system, then
in the non-interacting limit, we find that the 3rd and the 6th single
particle levels are flux independent. 
If we put 3 up and 3 down electron in the system, then in the
non-interacting limit as well as in the strong interacting
limit, 1 up and 1 down electrons are localized in the stub.
If we put 6 up electrons and 6 down electrons in the system,
then again in both limits, 2 up and 2 down electrons are
localized in the stub. In all these different systems,
the ratio of charge in the stub to the charge in the ring
is different. But, nevertheless, quite generally 
an integer amount of charge
gets localized in the stub. If we go away from the Fano regime,
then it is not possible to have an integer amount of charge
in the stub and hence the behavior of the system cannot be understood
in terms of the levels of the ring or in terms of the levels
of the stub.

It can be seen that in the strong interaction limit, in the presence of
NSDN, the charge density in the stub can be much higher 
(or lower) than the 
charge density in the ring. Charge in the stub increases as we increase the
repulsive interaction between the electrons and saturates when the
total charge in the stub becomes an integral number same as that
in the non-interacting limit. This is as if
there is an effective attraction between the electrons in the stub.
One may think that this effect is because the stub sites have a different
coordination number than the ring sites. For example site A in Fig. 1 (a)
has only one nearest neighbor while all sites in the ring has at least
two nearest neighbors. However, this effect 
happens for all possible parameters
that gives different ratio of total charge in the stub to the
total charge in the ring. Also, in Fig 1 (b) the sites in the bubble
are equivalent to the sites in the ring in regard to coordination number.
If we do similar calculation for Fig. 1 (b) 
as we did in Fig. 5, then we expect $<n_A>=<n_B>$ due to the
symmetry of the two sites $A$ and $B$. We also find that
that for the same interaction strength and same number of
ring sites as that in Fig. 5,
$<n_A>=<n_B>=0.491$, which is very close to the average
of the two solid lines in Fig. 5. This strongly suggests that coordination
number is not important. It is the non-symmetry dictated node
that makes the total charge in the stub or bubble similar to that
of the non-interacting system. As a result the total charge
in the stub or bubble tends to be an integer in spite of strong
repulsion. The combined effect of a NSDN and strong repulsive 
interaction is like an effective electron-electron attraction
in the stub or in the bubble. 

The clean interacting ring or a ring with a defect
is well understood in terms of
an effective Hamiltonian \cite{mat}, because of which we
can understand the magnetization of a ring of any arbitrary number of
sites and arbitrary number of electrons. In our case, after
electrons are localized in the stub, the stub behaves like
a static impurity. This gives us an easy way of understanding
ring stub systems that are beyond the reach of exact
diagonalization methods. When applying approximate methods,
this can provide a simple guideline for cross checking
results in some limits like the Fano limit.

Quantum mechanical study of coupled atoms date back to 1923
\cite{bor}. Analogous study has also been made using quantum
dots \cite{dyb}. The starting point is well separated quantum dots
or two dots with a large electrostatic barrier in between.
Then one gradually enhance the coupling between the dots
by decreasing the barrier or by bringing the dots close
to each other. Thus charge in each dot can be controlled
and made integer by physically decoupling the two dots.
The study of coupled geometries as presented
in this paper also seems to be an avenue for searching new
phenomenon and physics.
We show that we can have integral charge in the two
sub-systems not by physically decoupling them but
due to an internal mechanism.

\section{Conclusions}
 
The non-interacting wave function in coupled systems can have
NSDN. Interactions perturb the non-interacting wave function
and the NSDN is destroyed. But in the limit of strong
electron-electron interactions, again the NSDN comes into play
and create many counter-intuitive effects. Like the charge
fluctuation between the sub-systems is quenched. Also the
total charge in the stub or the bubble tends to the same
integral value as in the non-interacting case, defying the 
repulsive interaction between electrons. This is a
general feature whenever there is NSDN and does
not depend on charge distribution. Because of the
quenching of charge fluctuations between the ring and the stub,
the persistent currents in the ring can be understood
in terms of a ring decoupled from the stub. This can help us
understand larger systems for which numerical diagonalization is 
not possible.

\section{Acknowledgment}

We acknowledge the financial support from
the Academy of Finland and from the European Community
project ULTRA-1D (NMP4-CT-2003-505457)

\centerline{\bf Figure Captions}

\noindent Fig. 1. (a) A schematic diagram of a ring-stub system.
The system is made up of R sites in the ring and S sites in the wire
or stub. The sites are shown as big dots and the connection between them
is shown as lines. Sites in the stub are labeled A and B.
(b) A schematic diagram of a ring-bubble system.
Sites in the bubble are labeled A and B.

\noindent Fig. 2. Eigen energies of the single
particle states of the system in Fig. 1 (a),
as a function of $\alpha = 2 \pi \phi / \phi_0$,
in the non-interacting limit of $U=0$ and $V=0$. 
Here $\phi$ is the total flux through the ring and $\phi_0 =hc/e$.
The ring consists
of 9 sites and the stub consists of 2 sites. $E$ represents energy
and the unit of energy is $E_0=t$.

\noindent Fig. 3. Schematic diagrams of a stub (a) and
bubble (b) corresponding to Fig. 1, but now with open leads.

\noindent Fig.4. The system under consideration is in Fig. 1 (a) with
$R$=9 and $S$=2. There are 4 spin up electrons and 4 spin down electrons
in the system with $U$=2 and $V$=1. We show here spin up occupation
probability $<n_A>$ and $<n_B>$ 
of the two sites $A$ and $B$ in the stub as a function
of $\alpha$ where $\alpha=2 \pi \phi / \phi_0$. Here $\phi$ is
the total flux through the ring and $\phi_0 =hc/e$.

\noindent Fig. 5. Same as in Fig. 4, but $U$=9 and
$V$=5.  $<n_A>$ and $<n_B>$ are individually flux dependent but
the sum of them do not depend on flux.

\noindent Fig. 6. The first 6 many body levels
of a ring with a stub
as a function of $\alpha = 2 \pi \phi / \phi_0$. 
$\phi$ is
the total flux through the ring and $\phi_0 =hc/e$.
Here $R$=6, $S$=2, $U=10$ and $V$=8.
The unit of energy $E$ is $E_0=t$.

\noindent Fig. 7. Many body levels of a ring
with a potential impurity as a function of $\alpha = 2 \pi \phi / \phi_0$.
$\phi$ is
the total flux through the ring and $\phi_0 =hc/e$.
Here $R$=6, $S$=0, $U=10$ and $V$=8.
One of the sites has a site energy of $5t$.
The unit of energy $E$ is $E_0=t$.

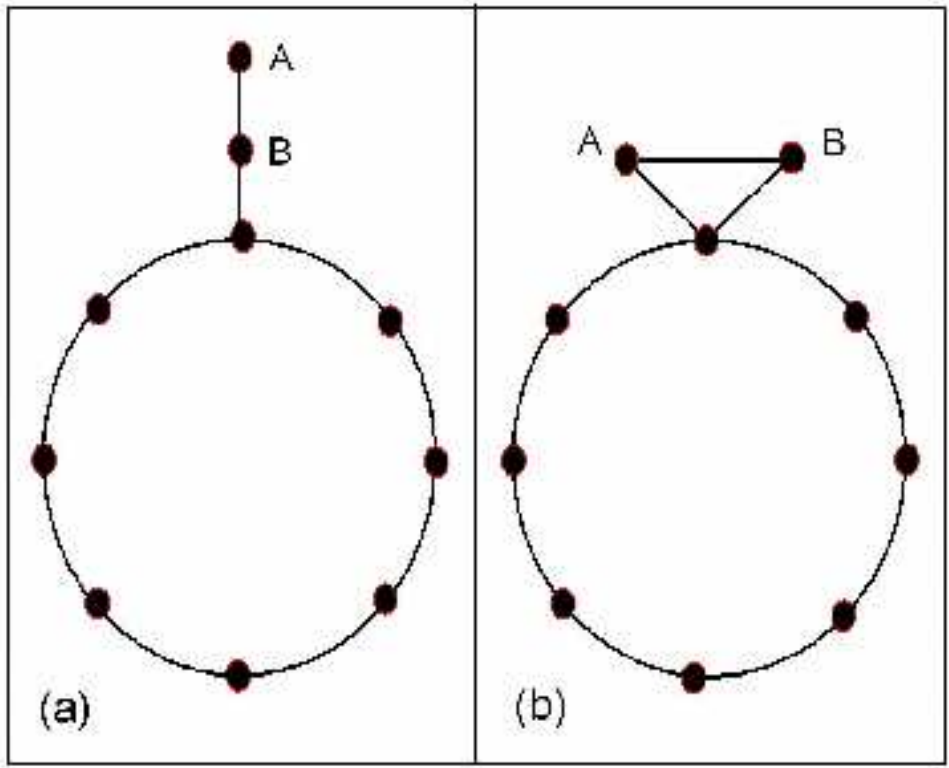
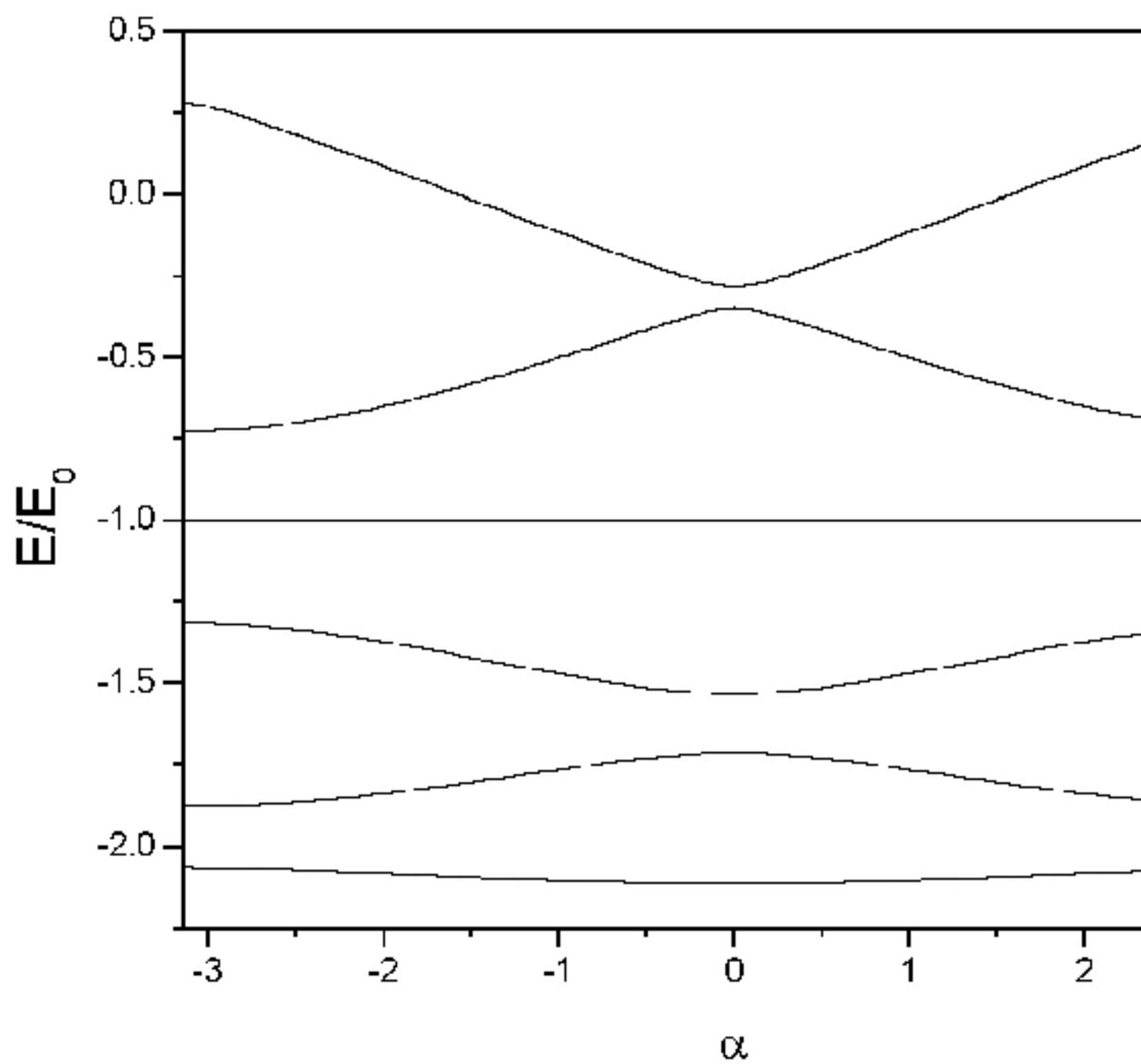
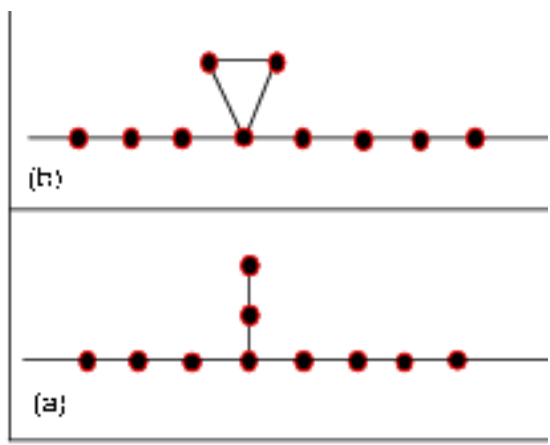
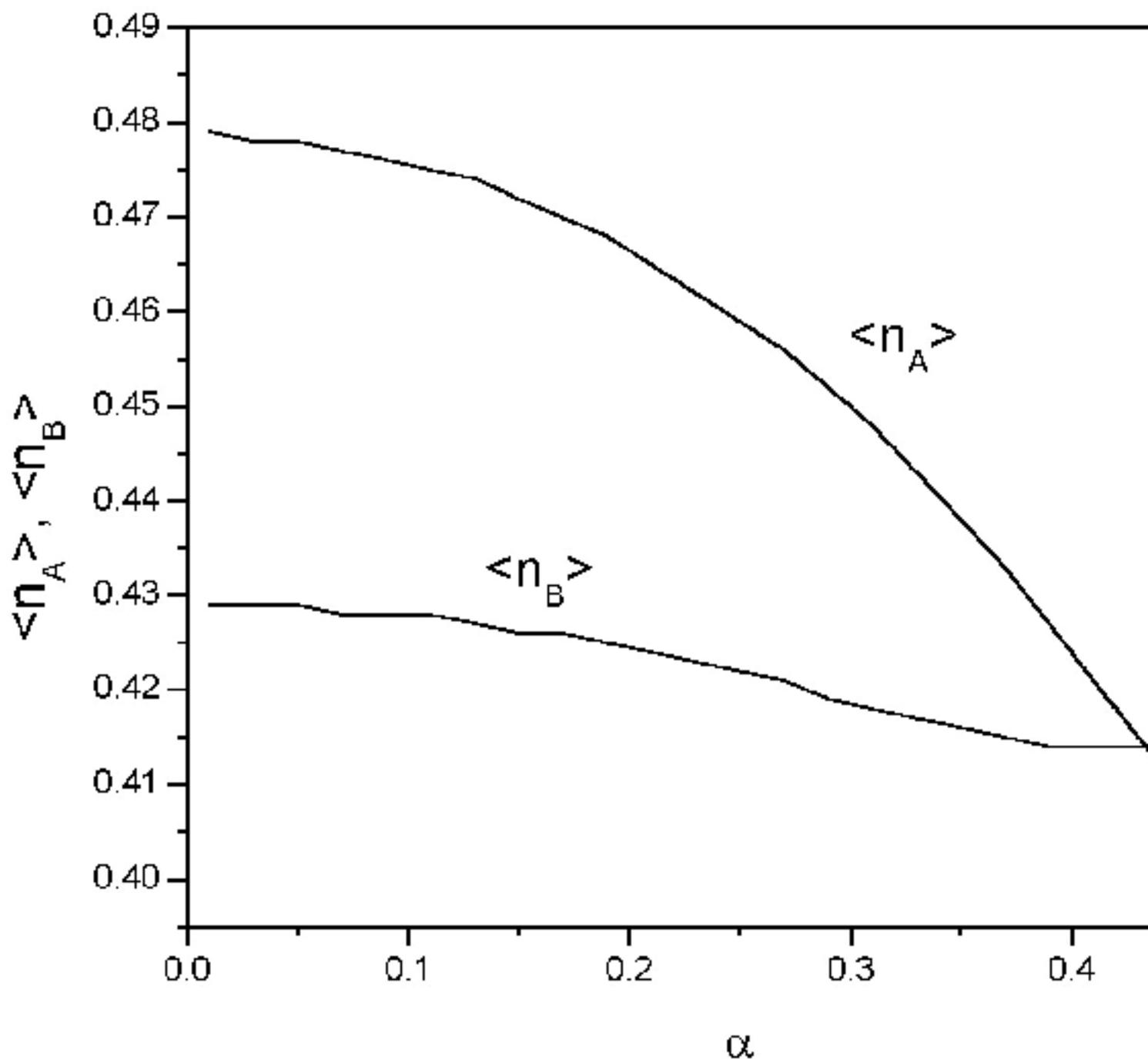
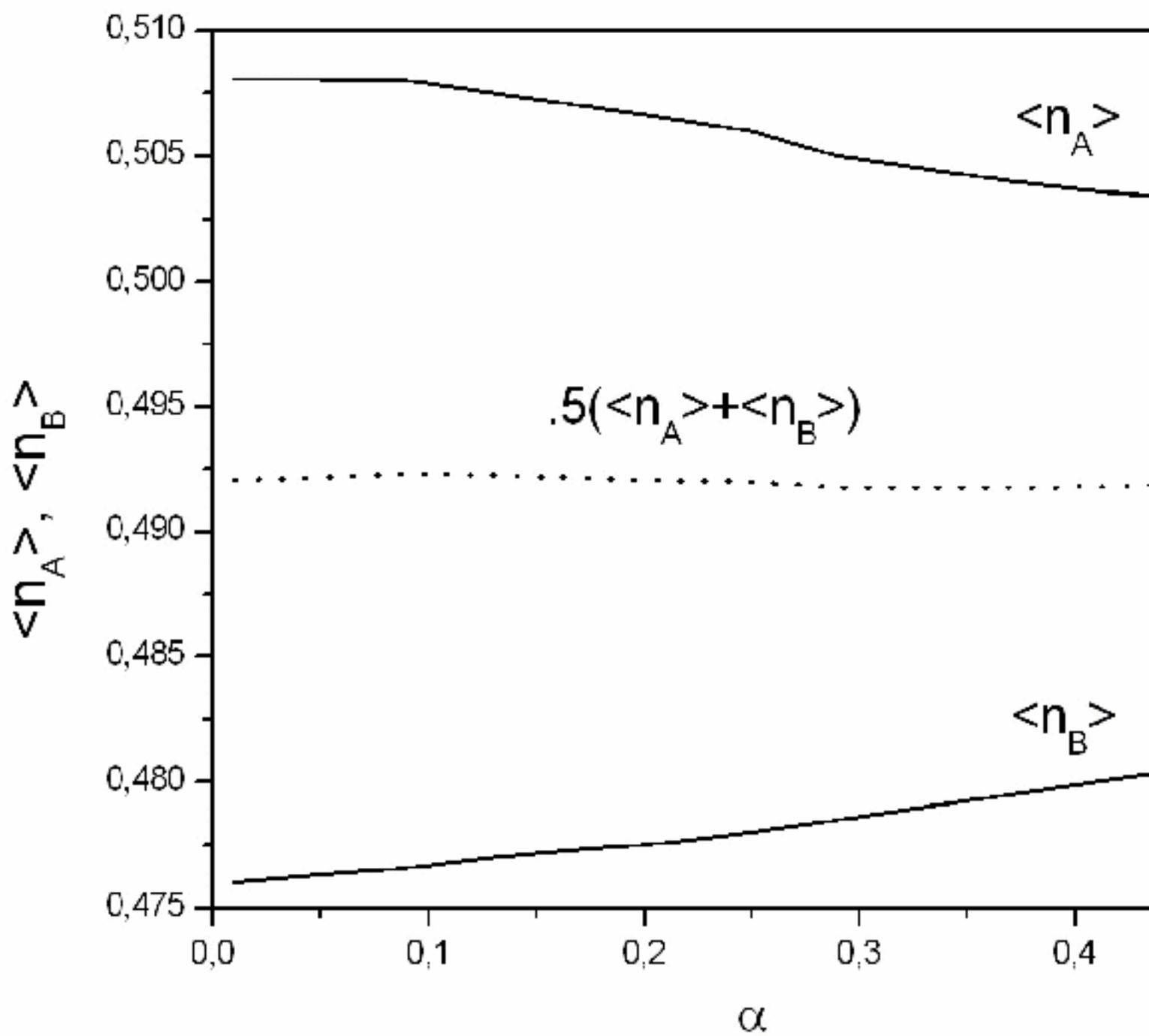
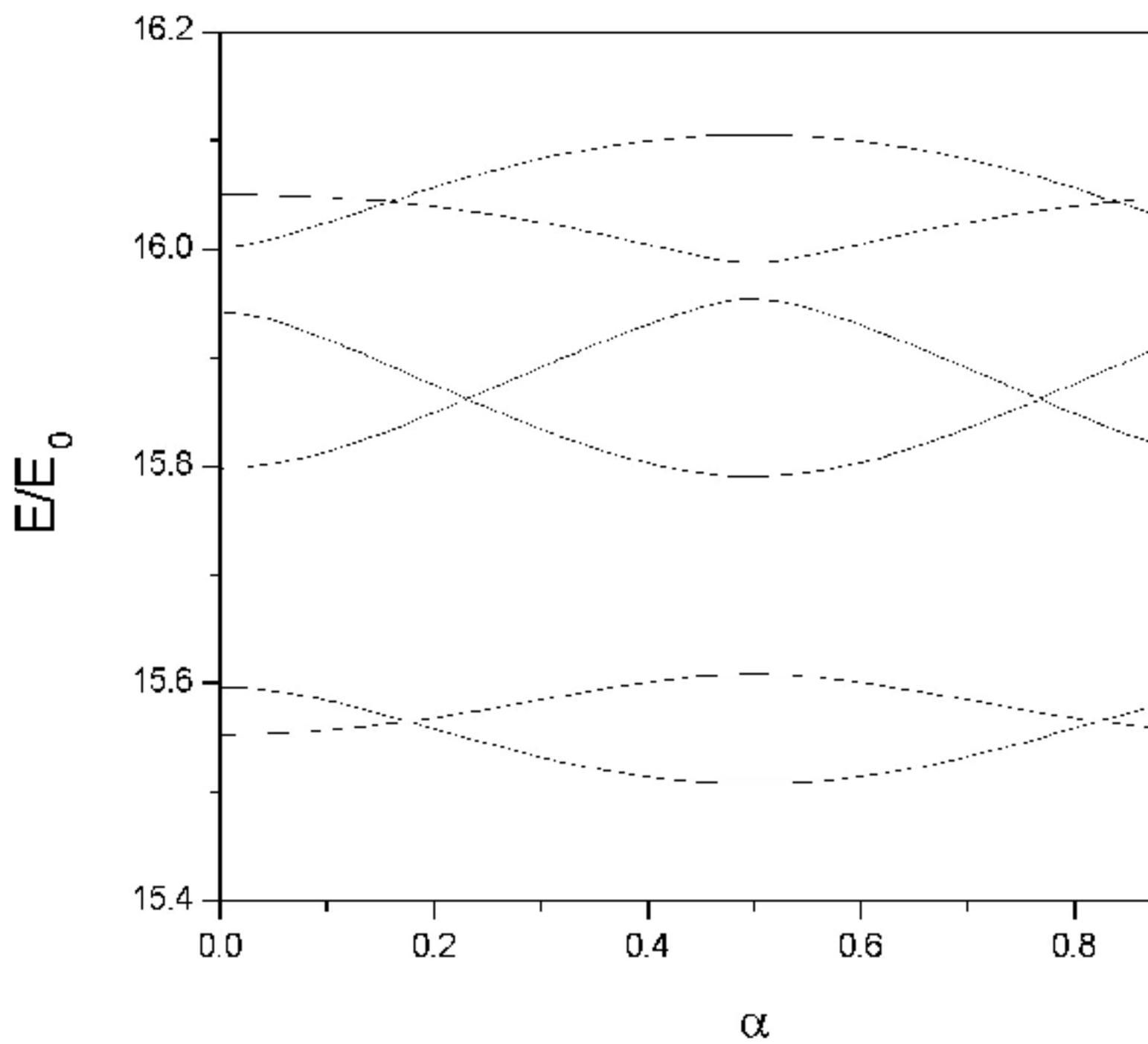
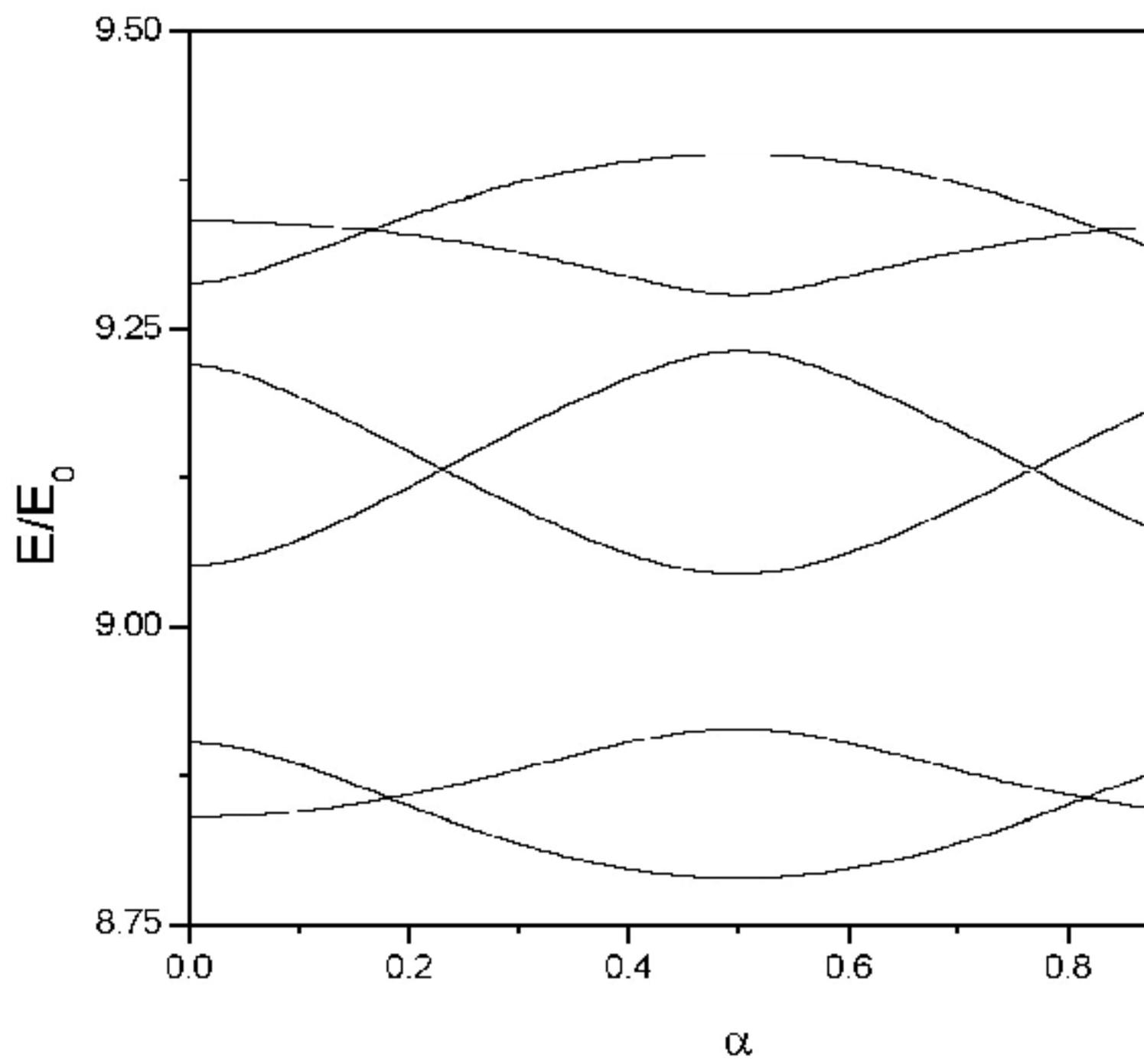
\end{document}